\documentstyle[aps,prl,epsfig,floats,twocolumn]{revtex}

\begin{document}
\title{Coulomb gap, Coulomb blockade, and dynamic activation energy in frustrated single-electron arrays}
\author{Daniel M. Kaplan, Victor A. Sverdlov, and Konstantin K. Likharev}
\address{
Department of Physics and Astronomy, State University of New York,
Stony Brook, New York 11794-3800 }

\date{\today}
\maketitle
\begin{abstract}
We have used modern supercomputer facilities to carry out extensive numerical simulations of statistical properties of 1D and 2D arrays of single-electron islands with random background charges, in the limit of small island self-capacitance. In particular, the spectrum of single-electron addition energies shows a clear Coulomb gap that, in 2D arrays, obeys the Efros-Shklovskii theory modified for the specific electron-electron interaction law. The Coulomb blockade threshold voltage statistics for 1D arrays is very broad, with r.m.s. width $\delta V_t$ growing as $\langle V_t \rangle \propto N^{1/2}$ with the array size $N$. On the contrary, in square 2D arrays of large size the distribution around $\langle V_t\rangle \propto N$ becomes relatively narrow $(\delta V_t/\langle V_t\rangle \propto 1/N)$, and the dc $I$-$V$ curves are virtually universal. At low voltages, the slope $G_0(T)$ of $I$-$V$ curves obeys the Arrhenius law. The corresponding activation energy $U_0$ grows only slowly with $N$ and is considerably lower than the formally calculated "lowest pass" energy $E_{max}$ of the potential profile, thus indicating the profile "softness".
\end{abstract}

PACS numbers: 73.23.Hk, 73.40.Rw, 85.35.Gv

\bigskip

\section{Introduction}

Arrays of small conducting islands separated by tunnel junctions are one of the simplest, and most important systems that exhibit the Coulomb blockade and the related effects of correlated single-electron tunneling - see, e.g., Refs. 1, 2. Properties of uniform arrays, at least within the framework of the "orthodox" theory of single-electron tunneling \cite{1,2}, for both 1D \cite{3,4,5,6} and 2D \cite{6,7,8,9} geometries, may be readily interpreted in terms of motion of single-electron solitons and anti-solitons with linear size (in terms of island number) $M \approx \max [1, (C/C_0)^{1/2}]$, where $C_0$ is the island self-capacitance, and $C$ the capacitance between the adjacent islands.

However, in practical arrays the apparently unavoidable charged impurities induce random shifts of the island electrochemical potentials, that are equivalent to random "background charges" $Q_j$ of the islands, spread uniformly within the range $[-e/2, +e/2]$ \cite{10} - for a discussion see, e.g., Sec. V.C of Ref. 11. The soliton language is inconvenient for the discussion of this case, and very little had been known about properties of such "completely frustrated" arrays. In the pioneering work \cite{12}, Middleton and Wingreen have shown that the average Coulomb blockade threshold voltage $V_t$ of frustrated 1D arrays with $C_0 \gg C$ (i.e., $ M \sim 1$) grows linearly with the array size $N$, while the r.m.s. width, $\delta V_t$, of its distribution  scales as $N^{1/2}$. Melsen {\it et al.} \cite{13} found that for 1D arrays with small island capacitance ($M \gg N$), $\langle V_t\rangle$ is proportional to $N^{1/2}$. Independently, Korotkov \cite{14} got the same result and also showed that in the former limit the relative width of the $V_t$ distribution, $\delta V_t/\langle V_t\rangle$, tends to a constant as $N \rightarrow \infty$. 

Even less was known about 2D arrays.  For the simplest case of large island capacitances $C_0 \gg C$, Middleton and Wingreen \cite{12} have found  that at $N \rightarrow \infty$, $\langle V_t\rangle \propto N$ and $\delta V_t/\langle V_t\rangle \propto  N^{-2/3}(\ln N)^{1/2}$. To our knowledge, all calculations for a finite soliton size ($C \sim C_0$) \cite{15,16} (as well as some calculations for 1D arrays \cite{17,18}) have been of illustrative character only.

The goal of this work was to establish the basic properties of the frustrated 1D and 2D arrays, with an emphasis on the statistics of their major parameters including not only $V_t$, but also the single-electron addition energy $E$ and the effective activation energy $U_0$ that determines the low-temperature linear conductance of the array. For the sake of simplicity we have studied the most interesting case of small island self-capacitance $(C_0 \ll C, C/N^2)$, i.e. of a large soliton size $(M \gg N)$. In contrast to such static parameters as $V_t$ and $E$, the calculation of $U_0$ requires a certain model of single-electron transport; for that we have used the "orthodox" theory of single-electron tunneling \cite{1}, that is qualitatively valid for metallic islands of a not very small size (practically, above  $\sim 1$nm - see Fig. 2 in Ref. 11).

	In order to reach reasonable accuracy of statistical calculations, this work has required substantial supercomputer resources that had become available for the academic solid state physics community only recently. Our algorithms were based on the well-known Monte Carlo code MOSES 1.2 that is available on the Web at http://hana.physics.sunysb.edu/set/software/index.html.

\section{Density of states and Coulomb gap}

	In the absence of voltage across a frustrated array, its main property is the statistics of single-electron addition energies $E$ of its islands. It might seem that the uniform distribution of $Q_j$ enables a straightforward calculation of the statistics, using the linearity of array electrostatics:

\begin{equation}
E_k = E_k^{(0)} + e \Sigma_j [C^{-1}]_{kj}Q_j,
\label{eqn:Art3.1}
\end{equation}
where $E_k$ is the single-electron addition energy of {\it k}-th island, i.e. the energy of a single-electron soliton centered at that island, $E_k^{(0)}=e^2[C^{-1}]_{kk}/2$ is its deterministic (and well known \cite{3,4,5,6,7}) value in the absence of background charges, while $C^{-1}$ is the reciprocal capacitance matrix of the array. However, due to the random character of $Q_j$, the initial distribution of these charges is generally unstable with respect to single-electron tunneling between the islands. Only after a series of such single-electron tunneling events, the system reaches a state with charges $Q_j^{(a)} = Q_j + en_j$, with some distribution \{$n_j$\} of additional charges, corresponding to the global minimum of electrostatic energy \cite{12}. 

\begin{figure}
\centerline{\hbox{
\psfig{figure=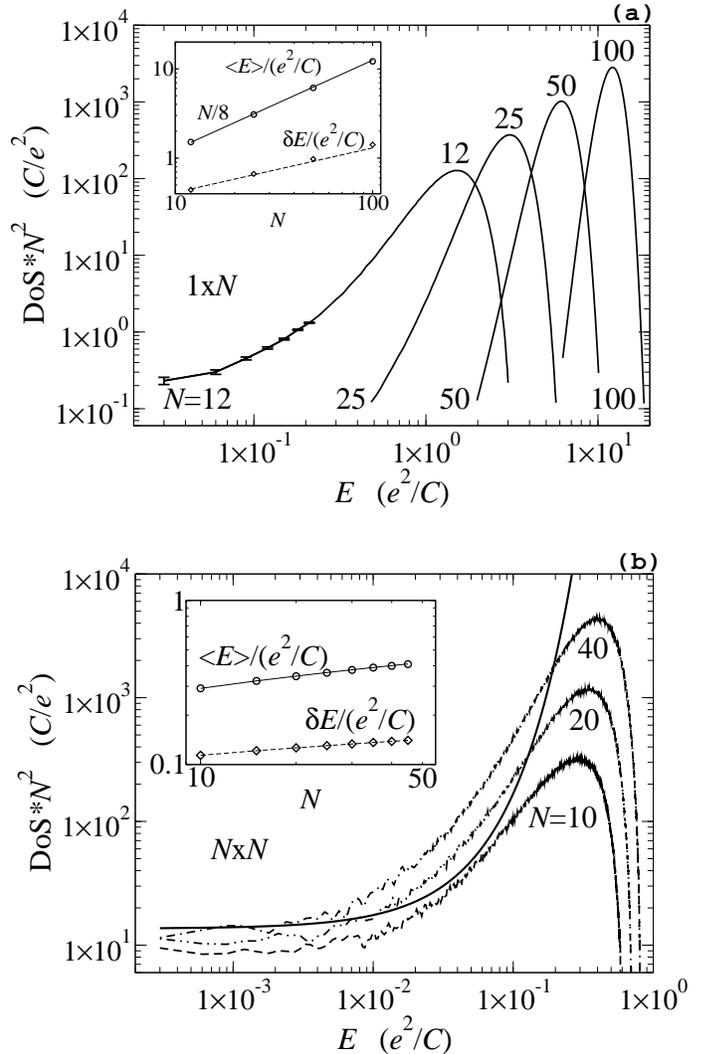,height=14cm,width=9.0cm}}}
\caption{Density of states DoS$\equiv dp/dE$ of (a) 1D and (b) 2D arrays as functions of the energy $E$ of single-electron addition to one their central islands, for several values of array size $N$. Solid line in Fig. \ref{fig:Ch4DoS}b shows the analytical result given by Eq. (\ref{eqn:Art3.1}). Insets show the average single-electron addition energy, and the r.m.s. width $\delta E$ of its distribution, as functions of $N$. Here, points are numerical results; solid lines show $E = E^{(0)}$ for arrays without frustration, while the dashed  the best fits for frustrated arrays: (a) $E^{(0)} = \langle E\rangle = Ne^2/8C$, $\delta E = 0.13 N^{1/2}e^2/C$; (b) $E^{(0)} = \langle E\rangle =  (e^2/4 \pi C) \ln(3.9N)$,  $\delta E = 0.019 (e^2/C) \ln(40N)$.}
\label{fig:Ch4DoS}
\end{figure}

	In experiment, such "annealing" happens automatically and is typically very fast (much faster than the typical experiment time). However, the faithful reproduction of this process in computer simulations requires special attention, because the simple relaxation within the framework of the orthodox theory at $T = 0$ would typically result in the system trapped in one of the metastable "Coulomb glass" states. Generally, the annealing may be carried out by increasing temperature $T$ (that affects the tunneling rates \cite{1}) and then gradually decreasing it to zero. We have found, however, that a faster annealing may be achieved by the following procedure carried out at $T = 0$: for a given island $k$, the single-electron addition energy $E_k$ is calculated. If the energy is negative (i.e., the electron addition from outside is energy-favorable), the electron is added ($n_k \rightarrow n_k + 1$), and the operation is repeated. If $E_k$ is positive, the system is allowed to relax due to tunneling between neighboring islands. Then the same procedure is carried out for the addition of a hole to the same island (energy $E'_k$). This process is repeated sequentially for all islands, until the addition of either an electron or a hole to any island is energy-unfavorable (all $E_k, E'_k \geq 0$). By definition, this is the annealed state of the system. Due to the electron/hole symmetry of single-electron tunneling within the orthodox theory \cite{1}, for the annealed state the probability distributions $p(E)$, $p(E')$ of  single-electron and single-hole excitations coincide. Because of the finite array size these distributions depend on the island position; however, they are virtually the same for a substantial number of the islands near the array center.

	Figures \ref{fig:Ch4DoS}a,b show the density of states DoS $= dp/dE$ for the central islands of, respectively, 1D arrays with ($N-1$) islands and 2D arrays with $(N-1) \times N$ islands for several values of array size $N$, while the insets show the average values of $E$ and r.m.s. deviations, $\delta E$, from those values (i.e., energy distribution width). A suppression of the density of states at $E \rightarrow 0$ is clearly visible. This is essentially the "Coulomb gap", first discussed by Efros and Shklovskii \cite{19} (see also Ref. \cite{20}). In the annealed state, electron transfer from island $k$ to island $j$ can only lead to an increase of the total energy of the system, i.e.

\begin{equation}
E_j + E'_k - U_{jk} > 0.
\label{eqn:Art3.2}
\end{equation}
where $U_{jk}=e^2[C^{-1}]_{kj}$ is the energy of electron-electron interaction. This requirement forbids a large part of configurations \{$n_j$\} and hence reduces the number of possible states, especially at small energies, just as in hopping systems \cite{19,20}. The only difference is quantitative: the electron-electron interaction law in the arrays is different from the simple unscreened Coulomb potential $U_{jk} = e^2/\epsilon |{\bf r}_j - {\bf r}_k|$ used for the description of hopping conductors. 

	For example, in 2D arrays of large size $N \gg 1$ (but still $N \ll M$) the interaction is close to 

\begin{equation}
U_{jk} = (e^2/2 \pi C) \ln(0.75N/|{\bf r}_j - {\bf r}_k|),
\label{eqn:Art3.3}
\end{equation}
for $1 \ll r_j ,r_k \ll N$, i.e. for islands close enough to the array center. Repeating the well-known arguments \cite{19,20} for this interaction law, we get the low-energy density of states

\begin{equation}
DoS(E) \approx 13.5(C/e^2N^2) \exp (25CE/e^2).
\label{eqn:Art3.4}
\end{equation}

This dependence (shown by solid line in Fig. \ref{fig:Ch4DoS}b) is in a very reasonable agreement with the numerical results, without any fitting parameters; in particular the predicted dependence DoS$(0) \propto N^{-2}$ is satisfied surprisingly well, given the approximate character of scaling arguments and of Eq. (\ref{eqn:Art3.3}). On the other hand, the similarity between the single-electron arrays and hopping conductors is not quite surprising, because the state of global energy minimum achieved at annealing does not depend on whether long electron hops are permitted (at hopping) or forbidden (at single-electron tunneling between adjacent islands).  

	For 1D arrays, however, the similar argumentation does not work at low excitation energies, because in this case the electron-electron interaction potential \cite{3} 

\begin{equation}
U_{jk} = (e^2/2C) (N/2 -|{\bf r}_j - {\bf r}_k|)
\label{eqn:Art3.5}
\end{equation}
does not approach zero within the range of validity ($1 \ll r_j ,r_k \ll N$) of Eq. (\ref{eqn:Art3.5}), and the usual arguments ignoring the system boundaries \cite{19,20} fail. Notice that in this case the Coulomb gap is almost "hard": for large $N$, DoS is exponentially low within a finite low-energy range of the order of $E^{(0)} \propto Ne^2/C$. 

	Returning to Fig. \ref{fig:Ch4DoS}, we should notice that the position of the DoS maximum, as well as the average single-electron addition energy $\langle E\rangle$, are both very close to that ($E^{(0)}$) for arrays without random background charge.  (For 1D case, $E^{(0)} = Ne^2/8C$ \cite{3}, while for 2D arrays, $E^{(0)} \approx (e^2/4 \pi C) \ln(\alpha N)$, with $\alpha = 3.9 \pm 0.1$). This means, in particular, that for 1D arrays even the ultimately large charge frustration does not change the single-electron addition energies considerably.

\section{DC I-V curves and Coulomb blockade}

\begin{figure}[tbp]
\centerline{\hbox{
\psfig{figure=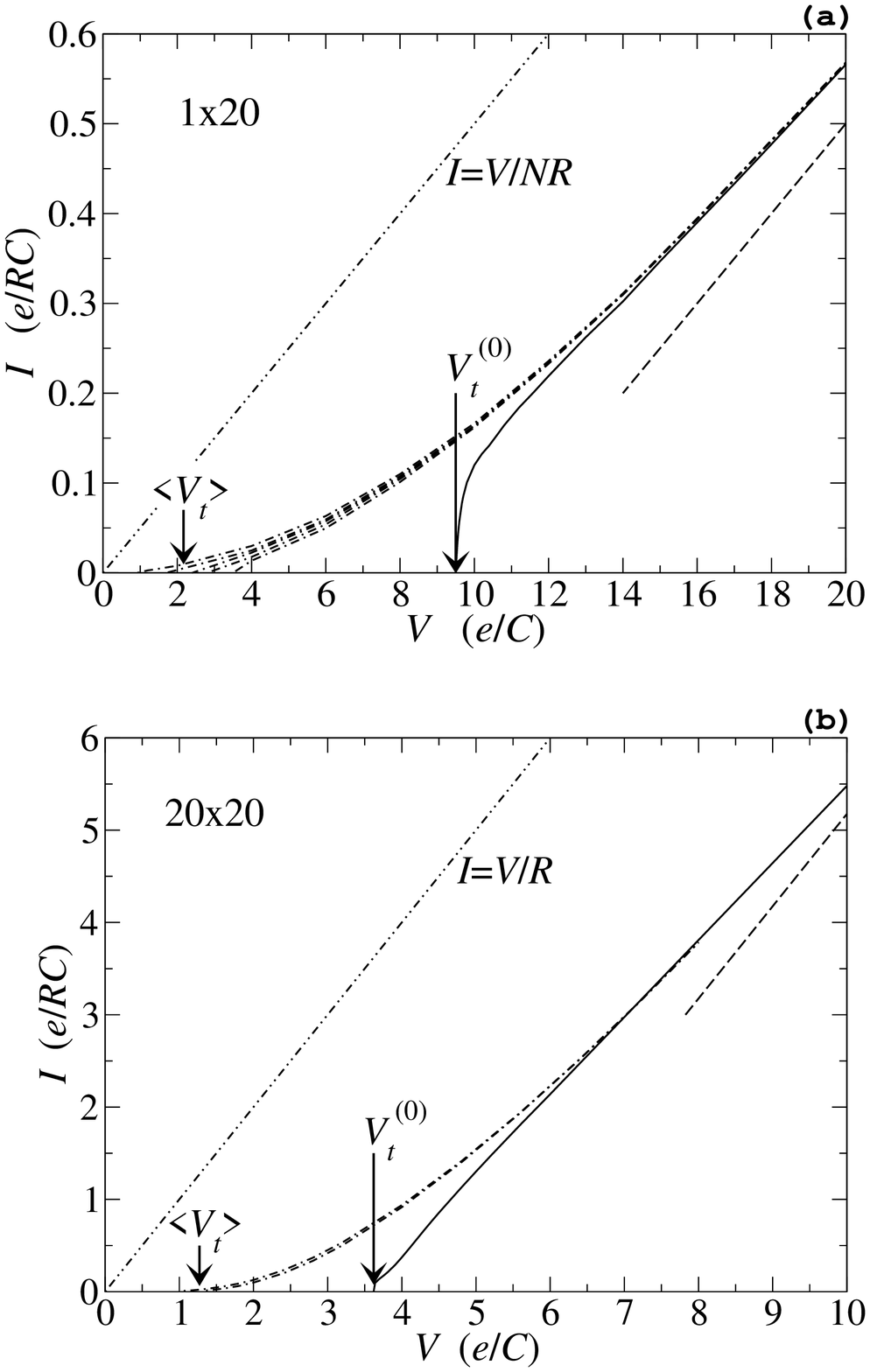,height=14cm,width=9.0cm}}}
\vspace{0.2cm}
\caption{DC $I$-$V$ curves of (a) 1D and (b) 2D arrays, both without background charge (solid lines) and with completely random background charge (dash-dotted lines). Dashed lines show the far asymptotes of the curves (a) 1D [3, 7], $V(I) \rightarrow  N [IR + $sgn$(I) e/2C]$ and (b) 2D $V(I) \rightarrow  [IR + $sgn$(I) Ne/4C]$, that are independent of the background charge. $R$ is the tunnel resistance of the single junction between the adjacent islands.}
\label{fig:Ch4IVlin}
\end{figure}

\begin{figure}[tbp]
\centerline{\hbox{
\psfig{figure=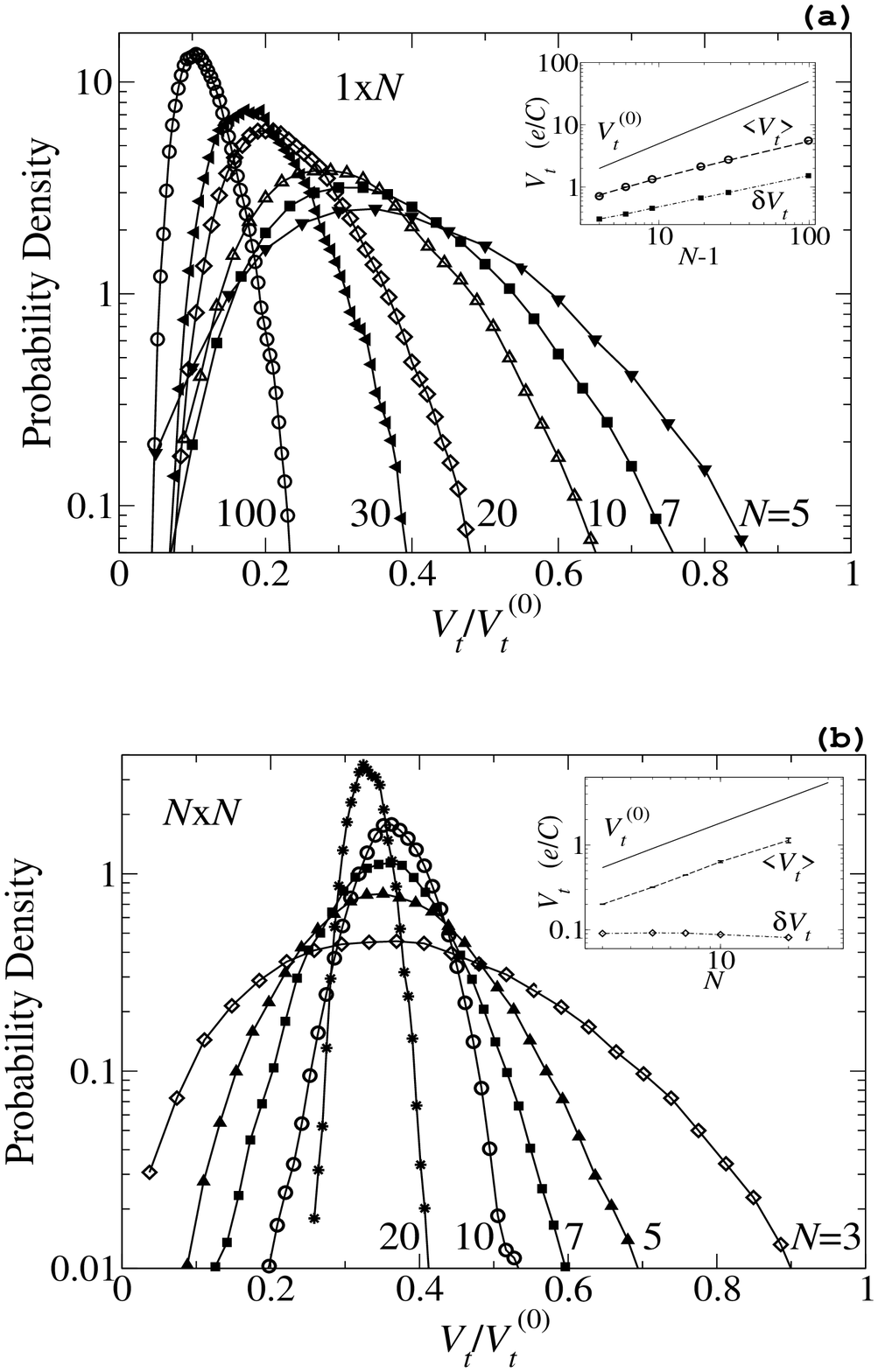,height=14cm,width=9.0cm}}}
\vspace{0.2cm}
\caption{Statistical distributions of the low-temperature threshold voltage $V_t$ for (a) 1D and (b) 2D arrays of various size. The insets show the values of $V_t$ without background charge (solid curves), and the average $V_t$ (dashed curves) and the r.m.s. width of their statistical distributions (dot-dashed curves) for frustrated arrays. In inset (a), the dashed line shows the phenomenological dependence $\langle V_t\rangle \approx 0.6 (N^{1/2} - 1)$ suggested in Ref. 13; the other lines are just guides for the eye.}
\label{fig:Ch4Vt}
\end{figure}

	Figures \ref{fig:Ch4IVlin}a,b show several dc $I$-$V$ curves of the frustrated arrays for $T=0$. One can see that for 2D arrays of large size the statistical variations of the curves are really small, and in this sense the curves are virtually universal. (They do not depend on whether the arrays had been annealed before the dc voltage application, because the motion of single-electron solitons at $V > V_t$ performs effective "dynamic annealing" of the systems.) It means that the Coulomb blockade threshold voltage $V_t$ (at which current vanishes if $T = 0$) also has a relatively narrow statistical distribution. This distribution is shown in Fig. \ref{fig:Ch4Vt}b; in contrast with the case  $C \ll C_0$ \cite{12}, its r.m.s. width, expressed in units $e/C$, is virtually constant while $\langle V_t\rangle$ grows as $N$, just like the threshold ($V_t^{(0)}$) for the case of zero background charge. (The ratio $\langle V_t\rangle/ V_t^{(0)}$ is close to 0.3.)

	On the contrary, in 1D arrays the statistical distribution of dc $I$-$V$ curves (Fig. \ref{fig:Ch4IVlin}a), and hence the threshold voltages (Fig. \ref{fig:Ch4Vt}a), is much broader: the $V_t$ distribution width grows with $N$ almost as fast as $\langle V_t\rangle$ (as ~$N^{1/2}$), so that the relative width $\delta V_t/\langle V_t\rangle$ decreases very slowly (from ~0.5 to ~0.4 between $N = 2$ to 100). Thus we have confirmed (with somewhat better accuracy) the prior 1D results \cite{13,14}.

\begin{figure}[tbp]
\centerline{\hbox{
\psfig{figure=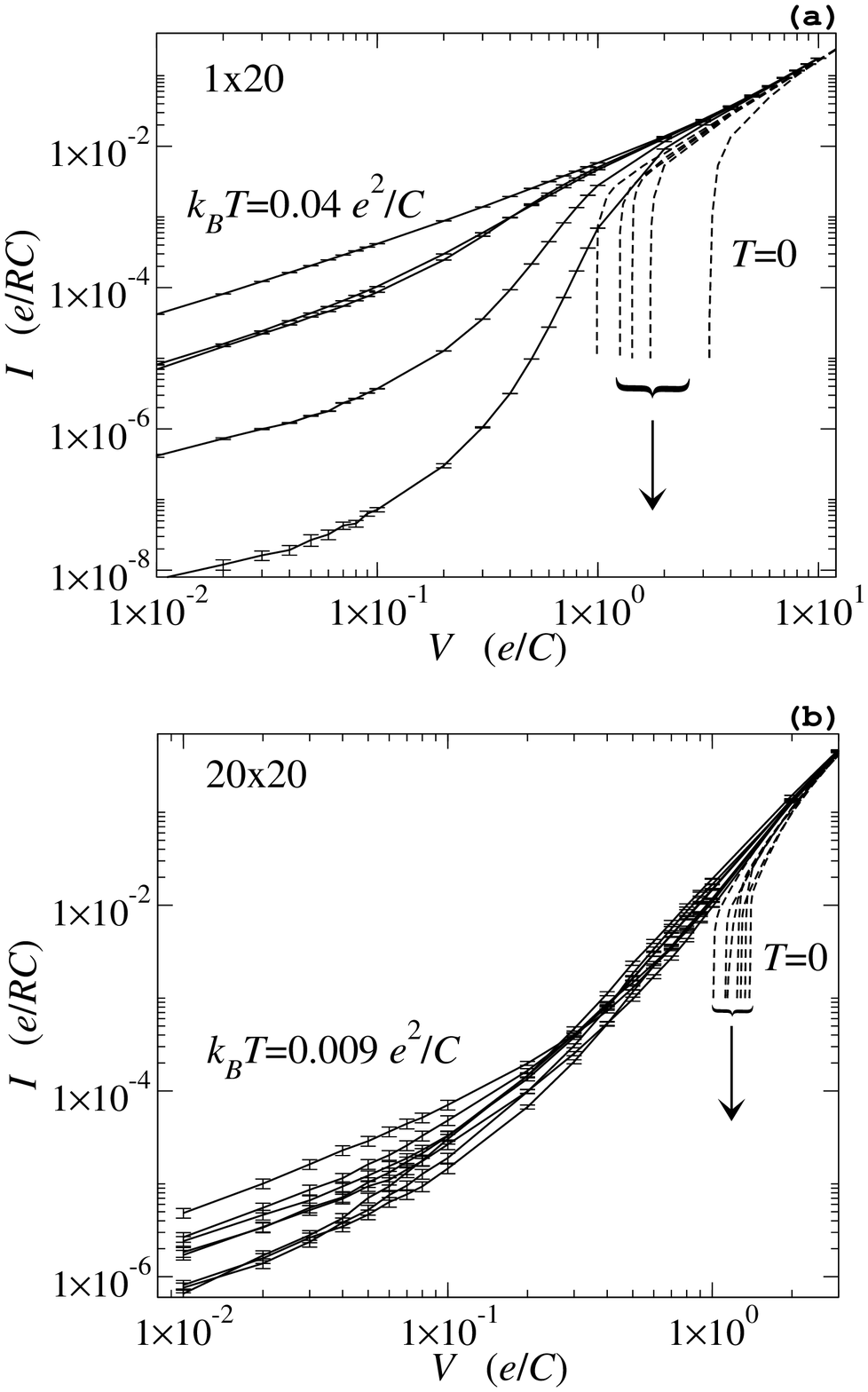,height=14cm,width=9.0cm}}}
\vspace{0.2cm}
\caption{Typical dc $I$-$V$ curves of several (a) 1D and (b) 2D arrays for low but finite temperatures, plotted in log-log  scale. The curves for $T = 0$ are shown with dashed lines for comparison.) The plots clearly show the linear conductance in the arrays at low voltages.}
\label{fig:Ch4IVlog}
\end{figure}

	As a by-product, we have found that, just like in the opposite case $C \ll C_0$ \cite{12}, the shape of dc $I$-$V$ curves at $V \gtrsim V_t$ is close to a power law $I \propto (V - V_t)^{\zeta}$, with $\zeta = 1.0 \pm 0.1$ in 1D and $\zeta = 1.7 \pm 0.1$ in 2D.

\section{Linear conductance and dynamic activation energy}

\begin{figure}[tbp]
\centerline{\hbox{
\psfig{figure=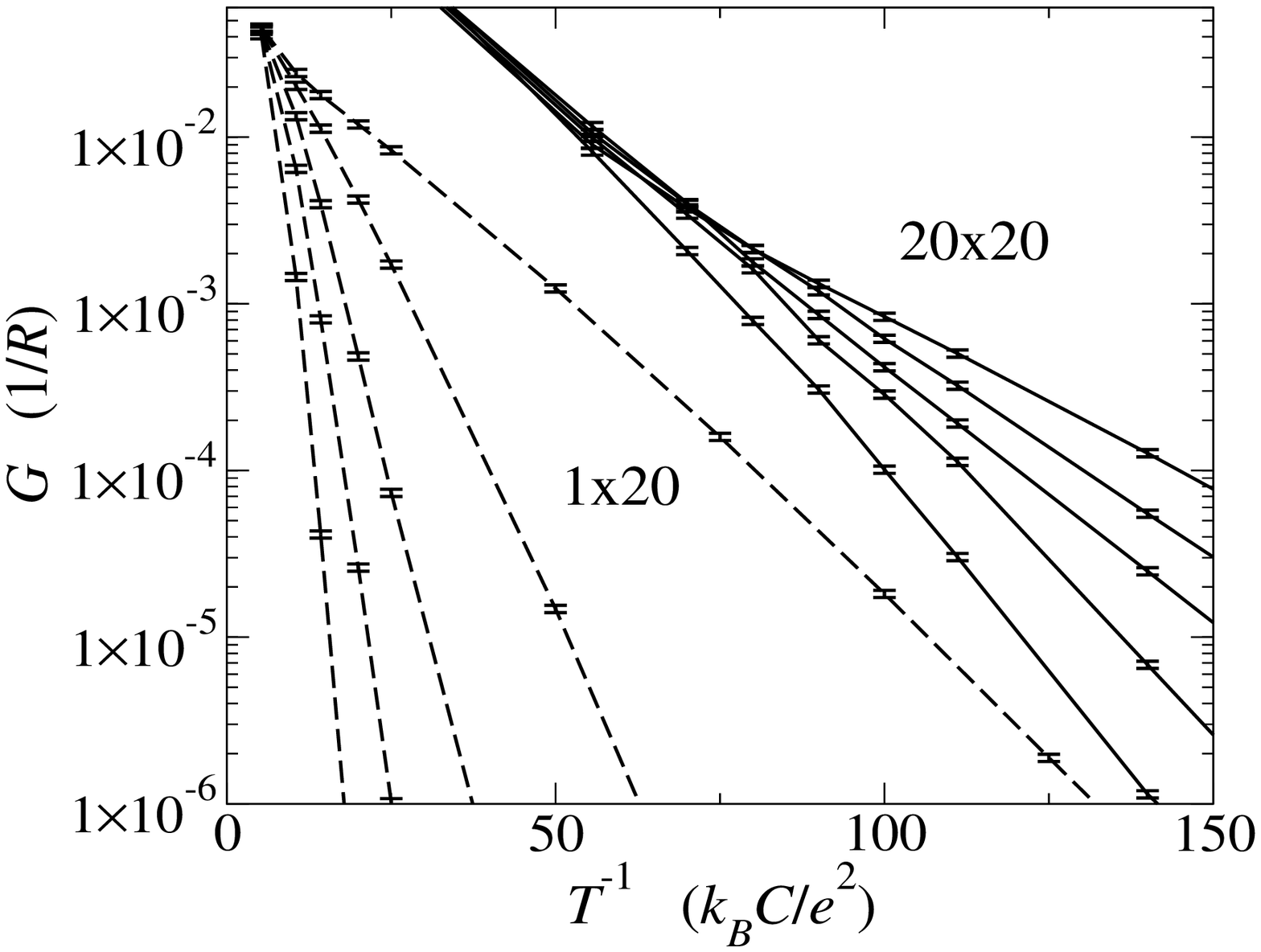,width=9.0cm}}}
\vspace{0.2cm}
\caption{The low-voltage conductance of 1D (dashed lines) and 2D (solid lines) arrays as a function of inverse temperature, for several random distributions of the background charge. At low temperatures the plots are linear, indicative of the Arrhenius law: $G \propto \exp(-U_0/k_BT).$}
\label{fig:Ch4GvsTinv}
\end{figure}

	Even small thermal fluctuations lift the Coulomb blockade \cite{1,2} and ensure a low but finite conductance $G \equiv I/V$ of the array at low applied voltages $|V| \ll V_t$. Figure \ref{fig:Ch4IVlog} shows that in frustrated arrays this conductance is also random, with 1D arrays again much more irreproducible than square 2D arrays of the same length. Figure \ref{fig:Ch4GvsTinv} shows that, in contrast to systems with variable-range hopping \cite{20}, at low temperatures the conductance follows the Arrhenius law

\begin{equation}
G \propto\ exp(-U_0/k_BT),
\label{eqn:Art3.6}
\end{equation}
with the dynamic activation energy $U_0$ depending on the random background charge distribution.

\begin{figure}[tbp]
\centerline{\hbox{
\psfig{figure=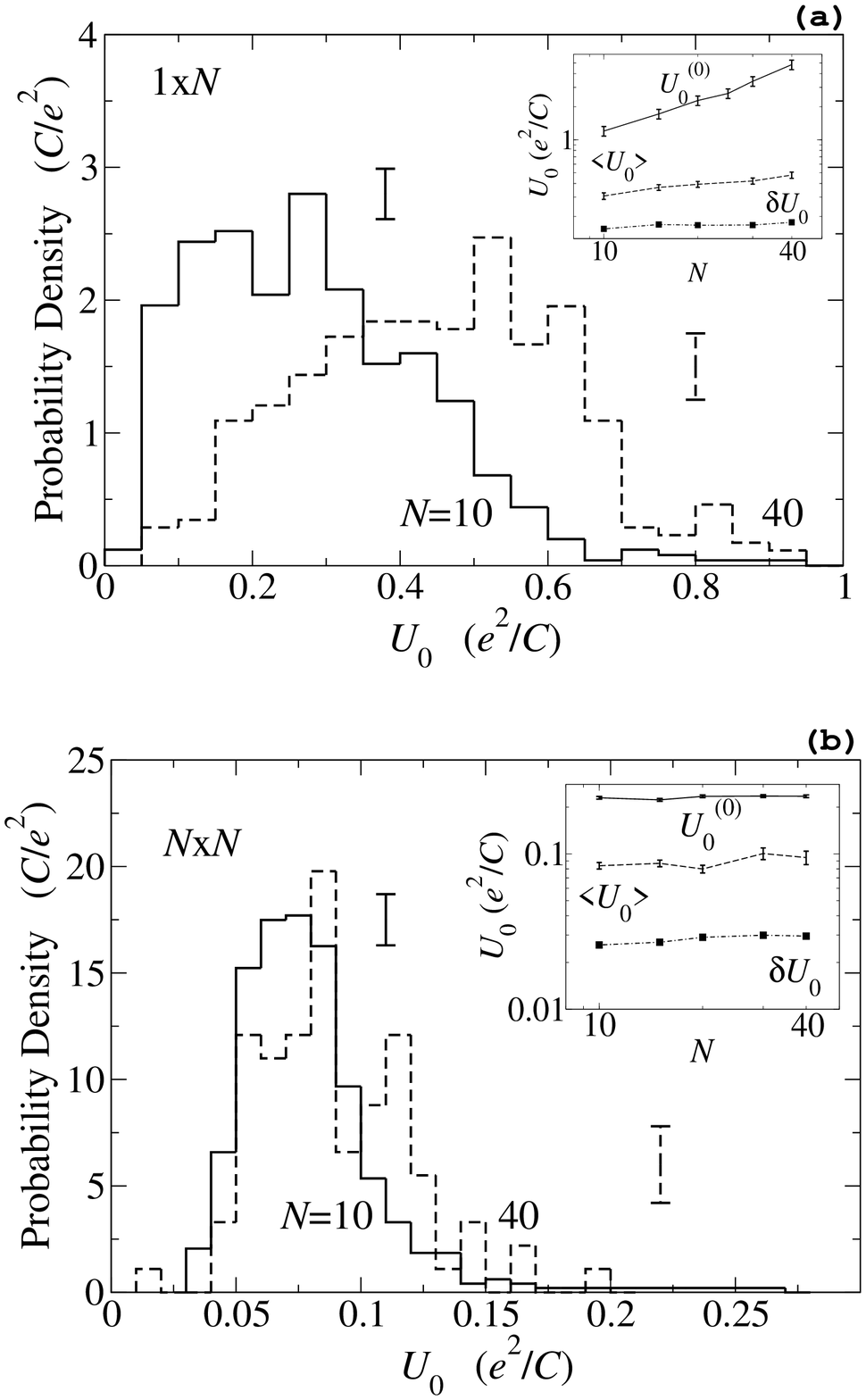,height=14cm,width=8.0cm}}}
\vspace{0.2cm}
\caption{Histograms of the dynamic activation energy $U_0$ for (a) 1D and (b) 2D frustrated arrays. (For clarity, they are shown only for the largest and smallest values of $N$ studied in this work.) The insets show the values of $U_0$ without background charge (solid curves), and the average $U_0$ (dashed curves) and the r.m.s. width of their statistical distributions (dot-dashed curves) for frustrated arrays. The calculated values are shown by dots; curves are only guides for the eye.}
\label{fig:Ch4Uo}
\end{figure}

	Figures \ref{fig:Ch4Uo}a,b show typical histograms of the statistical distribution of $U_0$, while insets show the dependence of its first two momenta on the array size. In 2D arrays (Fig. \ref{fig:Ch4Uo}b), neither of the dependences is very strong: $\langle U_0\rangle \approx 0.09 e^2/C$, $\delta U_0 \approx 0.03 e^2/C$. Notice that in 2D arrays without frustration ($Q_k \equiv 0$), the activation energy is also virtually independent of $N$: $U_0^{(0)} \approx 0.23 e^2/C$. On the contrary, in 1D arrays the average activation energy grows, albeit weakly, as $N$ is increased (Fig. \ref{fig:Ch4Uo}a).  It remains, however, much lower than its level $\langle U_0\rangle \approx E^{(0)} = Ne^2/8C$ in arrays without frustration. 

	It is also important to notice that the potential profile of strongly frustrated single-electron arrays is "soft", in the following sense. If we plot the single-electron addition energy of an annealed 1D array with $V = 0$ as a function of the island number $k$, we get a potential profile $E_k$. It is tempting to assume that the activation energy $U_0$ is just the maximum, $E_{max}$, of this profile, because in order to carry an electron through the array at small $V$, thermal fluctuations should overcome this threshold. (For 2D case, $E_{max}$ is determined as the lowest maximum of all paths connecting the opposite electrodes.) However, our numerical experiments have shown that this assumption is not true: $U_0$ is always lower than $E_{max}$ in any arrays (especially those with frustration). The simple interpretation of this fact is that when thermal fluctuations induce the transfer of an addition electron over the potential profile of a frustrated array, its Coulomb field causes shifts of the neighboring charges and thus deforms the profile, reducing its maximum.

	One more important fact is that in frustrated arrays of any dimensionality the statistical distribution of $U_0$ is very broad: $\delta U_0/\langle U_0\rangle \propto  0.2 - 0.3$ for $10 < N < 40$.

\section{Comparison with experiment}

	Until recently, most experimental studies have been carried out on systems with random island size and location - see, e.g., Refs. 21-23. In this case, the tunnel conductance between islands has exponentially broad statistical distribution, and this randomness dominates that created by the background charge frustration. However, there is a growing number of  experiments \cite{25,26,27,28,29} with 2D arrays of virtually similar and well-ordered islands, e.g. gold \cite{25,27,28,29} or cobalt \cite{26} clusters separated by apparently similar organic shells. (One can add the earlier work \cite{24}, where 1D and 2D arrays have been fabricated by nanolithography.) For such arrays, the charge frustration should determine the property statistics, and the comparison with our results might be meaningful. Unfortunately, for all these arrays the ratio $C_0/C$ is of the order of one, excluding the possibility of direct quantitative comparison of our results with the experimental data reported in these publications. (Our future plans include the extension of our calculations to this case.)

	However, several features of the experimental results are in a qualitative agreement with our calculations. In particular,

	- for large 2D arrays, the dc $I$-$V$ curves are virtually reproducible (in particular, $\delta V_t \ll \langle V_t\rangle $);

	- the general shape of these curves is rather close to that shown in Fig. \ref{fig:Ch4IVlin}b;

	- the linear conductance does obey the Arrhenius law (6) at low temperatures. 

\section{Discussion}

	The results discussed above show that the average values and distribution widths of activation energy $U_0$, the Coulomb gap, and the Coulomb blockade threshold $V_t$ of charge-frustrated arrays substantially differ from each other, so that all these three notions should be clearly distinguished. (This has not always been done in earlier publications.) 

	These results have significant practical implications. Two important applications for multi-junction arrays (sometimes called "multiple tunnel junctions", or MTJ) in future single-electron circuits have been suggested:

	(i) as a replacement for the double junction in single-electron transistors, in order to raise the single-electron addition energy and hence increase the operation temperature range \cite{21,30}; and

	(ii) as a circuit element with "sub-electron" (quasi-continuous) charge transfer, capable of leaking to ground the random background charge of single-electron islands \cite{11,31}.

	For application (i), the Coulomb blockade threshold of the array should be well reproducible (device-to-device), and controllable by voltage $V_g$ applied to an external gate electrode. Figure \ref{fig:Ch4Vt}b shows that in frustrated 2D (but not 1D!) arrays with large N the former condition is actually fulfilled ($\delta V_t \ll \langle V_t\rangle$); however, the second condition cannot be met. In fact, the action of external gate is always equivalent \cite{1} to insertion of an additional charge $(Q_k)_g \propto V_g$ into each island of the array. If the array is completely frustrated, its random background charges (mod e) are uniformly distributed within the interval $[-e/2, +e/2]$, and the addition of additional charges $(Q_k)_g$ does not change this statistics, and hence $\langle V_t\rangle$. 

	The minimum requirement for application (ii) is to have Coulomb blockade deeply suppressed \cite{31}: $V_t \ll e/C_e$, where $C_e$ is the effective capacitance of the island the array would shunt. Figure \ref{fig:Ch4Vt} shows, however, that for both 1D and 2D arrays with $N \gg 1$ (this condition is necessary for reproducibility, i.e. for having $\delta V_t \ll \langle V_t\rangle$) the average value of $V_t$ remains much larger than $e/C$. This means that capacitance $C$ (and consequently the island size) of the array islands should be much larger than those of the island the array should serve. For practical integrated circuits this is hardly a good solution. (For discrete devices and fundamental experiments, however, the shunting idea may work.)

	Notice that we have explored arrays with ultimately low self-capacitance $C_0$ of array islands. Elementary arguments show that for both applications discussed above the self-capacitance $C_0$ would have the detrimental effect; hence our negative conclusion cannot be affected by this limitation.

	Useful comments by A. Korotkov are greatfully acknowledged. The work was supported in part by the Engineering Physics Program of the Office of Basic Energy Sciences at the Department of Energy and by the Semiconductor Research Corporation. We also acknowledge the use of following supercomputer resources: SBU's cluster {\it Njal} (purchase and installation funded by DoD's DURINT program), Oak Ridge National Laboratory's IBM SP computer {\it Eagle} (funded by the Department of Energy's Office of Science and Energy Efficiency program), and also two IBM SP systems: {\it Tempest} at Maui High Performance Computing Center and {\it Habu} at NAVO Shared Resource Center's (computer time granted by DOD's High Performance Computing Modernization program).


\begin{references}
\bibitem{1} D. V. Averin and K. K. Likharev, in Mesoscopic Phenomena in Solids, ed. by B. Altshuler et al. (Elsevier, Amsterdam, 1991), p. 173.
\bibitem{2} Single Charge Tunneling, ed. by H. Grabert and H. Devoret (Plenum, New York, 1992).
\bibitem{3} N. S. Bakhvalov, G. S. Kazacha, K. K. Likharev, and S. I. Serdyukova, Sov. Phys. JETP {\bf 68}, 581 (1989). 
\bibitem{4} P. Delsing, in Ref. 2, p. 249.
\bibitem{5} A. N. Korotkov, Phys. Rev. B {\bf 50}, 17674 (1994).
\bibitem{6} J. Mooij, B. J. Van Wees, L. J. Geerligs, M. Peters, R. Fazio, and G. Sch\"on, Phys. Rev. Lett. {\bf 65}, 645 (1990).
\bibitem{7} N. S. Bakhvalov, G. S. Kazacha, K. K. Likharev, and S. I. Serdyukova, Physica B {\bf 64}, 173 (2001).
\bibitem{8} J. Mooij and G. Sch\"on, in Ref. 2, p. 275.
\bibitem{9} V. A. Sverdlov, D. M. Kaplan, A. N. Korotkov, and K. K. Likharev, Phys. Rev. B {\bf 64}, 041302 (2001).
\bibitem{10} The background charges may be defined modulo $e$ because of the unavoidable annealing of the system - see below.   
\bibitem{11} K. K. Likharev, Proc. IEEE {\bf 87}, 606 (1999).
\bibitem{12} A. A. Middleton and N. S. Wingreen, Phys. Rev. Lett. {\bf 71}, 3198 (1993).
\bibitem{13} J. A. Melsen, U. Hanke, H.-O. M\"uller, and K. A. Chao, Phys. Rev. B  {\bf 55}, 10638 (1997).
\bibitem{14} A. N. Korotkov, unpublished (1995), see Fig. 13 in Ref. 11.
\bibitem{15} H.-O. M\"uller, K. Katayama, and H. Mizuta, J. Appl. Phys. {\bf 84}, 5603 (1998).
\bibitem{16} H.-O. M\"uller, D. A. Williams, and H. Mizuta, Jpn. J. Appl. Phys. (part 2) {\bf 39}, L723 (2000).
\bibitem{17} J. Johansson and D. B. Haviland, Phys. Rev. B {\bf 63}, 014201 (2000).
\bibitem{18} V. L. Nguyen, T. D. Nguyen, and H. N. Nguyen, Phys. Lett. A {\bf 291}, 150 (2001).
\bibitem{19} A. L. Efros and B. I. Shklovskii, J. Phys. C {\bf 8}, L49 (1975).
\bibitem{20} B. I. Shklovskii and A. L. Efros, Electronic Properties of Doped Semiconductors (Springer, Berlin, 1984).
\bibitem{21} W. Chen, H. Ahmed, and K. Nakazato, Appl. Phys. Lett. {\bf 66}, 3383 (1995).
\bibitem{22} K. Yano {\it et al.}, Proc. IEEE {\bf 87}, 633 (1999).
\bibitem{23} A. S. Cordan, Y. Leroy, A. Goltzene, A. Pépin, C. Vieu, M. Mejias, and H. Launois, J. Appl. Phys. {\bf 87}, 345 (2000).
\bibitem{24} C. Kurdak, A. J. Rimberg, T. R. Ho, and J. Clarke, Phys. Rev. B. {\bf 57}, R6842 (1998).
\bibitem{25} R. P. Anders {\it et al.}, Science {\bf 273}, 1690 (1996).
\bibitem{26} C. T. Black, C. B. Murray, R. L. Sandstrom, and S. Sun, Science {\bf 290}, 1131 (2000).
\bibitem{27} R. Parthasarathy, X.-M. Lin, and H. M. Jaeger, Phys. Rev. Lett. {\bf 87}, 186807 (2001).
\bibitem{28} M. G. Ancona, W. Kruppa, R. W. Rendell, A. W. Snow, D. Park, and J. B. Boss, Phys. Rev. B {\bf 64}, 033408 (2001).
\bibitem{29} R. Parthasarathy, X.-M. Lin, K. Elteto, T. F. Rosenbaum, and H. M. Jaeger {\it et al.}, "Finite Temperature Electron Transport in Metal Nanocrystal Arrays", arXiv preprint cond-mat/0302348 (2003).
\bibitem{30} R. H. Chen and K. K. Likharev, Appl. Phys. Lett. {\bf 72}, 61 (1998).
\bibitem{31} K. A. Matsuoka and K. K. Likharev, Phys. Rev. B {\bf 57}, 15613 (1998).
\end{references}
\end{document}